\documentclass[acmsmall]{acmart}

\AtBeginDocument{%
  \providecommand\BibTeX{{%
    \normalfont B\b16kern-0.5em{\scshape i\kern-0.25em b}\kern-0.8em\TeX}}}

\usepackage{booktabs} 
\usepackage[utf8]{inputenc}
\usepackage[]{algorithm2e}
\usepackage{graphicx}

\begin{document}

\title{Towards automated verification of multi-party consensus protocols}

\author{Ivan Fedotov}
\affiliation{%
  \institution{Moscow Institute of Physics and Technology}
  \country{Russia}
  \streetaddress{Institutskiy Pereulok, 9}
  \city{Dolgoprudny} 
  \postcode{141701}
}
\email{ivan.fedotov@phystech.edu}

\author{Anton Khritankov}
\affiliation{%
  \institution{Moscow Institute of Physics and Technology}
  \country{Russia}
  \streetaddress{Institutskiy Pereulok, 9}
  \city{Dolgoprudny} 
  \postcode{141701}
}
\email{anton.khritankov@phystech.edu}

\author{Artem Barger}
\affiliation{%
  \institution{Idea Blockchain Competence Center}
  \country{Israel}
  \city{Haifa} 
}
\email{artem@bargr.net}

\begin{abstract}
 Blockchain technology and related frameworks have recently received extensive attention. Blockchain systems use multi-party consensus protocols to reach agreements on transactions. Hyperledger Fabric framework exposes a multi-party consensus, based on endorsement policy protocol, to reach a consensus on a transaction. In this paper, we define a problem of verification of a blockchain multi-party consensus with probabilistic properties. Further, we propose a verification technique of endorsement policies using statistical model checking and hypothesis testing. We analyze several aspects of the policies, including the ability to assign weights to organizations and the refusal probabilities of organizations. We demonstrate on experiments the work of our verification technique and how one can use experimental results to make the model satisfiable the specification. One can use our technique to design enterprise applications with the Hyperledger Fabric framework.
\end{abstract}

\ccsdesc[300]{Software and its engineering~Formal software verification}
\ccsdesc[300]{Software and its engineering~Software design engineering}
\ccsdesc[300]{Information systems~Enterprise applications}

\keywords{Verification, blockchain, consensus protocols, statistical model checking}

\maketitle

\section{Introduction}
Multi-party consensus protocols nowadays have widespread use in different application domains: logistic area, internet of things, financial services, and others. In blockchain applications, multi-party consensus protocol exposes the necessity to reach a consensus between untrusted parties which need to agree on changes in the system without the help of a third party.  

Since blockchain technology has become popular just recently, frameworks and tools for testing and verification of blockchain systems are still under development. Vulnerabilities in blockchain applications and particularly in multi-party protocols of Hyperledger Fabric can lead to significant losses \cite{b25}.

In our work, we address a statistical model checking of multi-party consensus protocols. Also, we show that one can apply our approach to verify policies in the Hyperledger Fabric framework (HLF). In particular, this paper makes the following contributions:
\begin{enumerate}
   \item States the verification problem for endorsement policies
   \item Provides algorithms to build a model and specification for consensus protocols
   \item Demonstrates usage of statistical model checking technique to verify consensus protocols
\end{enumerate}

The remainder of the paper is organized as follows.  Section \ref{sec:II} states the problem and outlines techniques that we used to solve it. In section \ref{sec:III} we provide related studies on multi-party consensus protocols verification. In section \ref{sec:IV} we explain in detail our algorithms to build a model and specifications from the properties of the system. We carry out experiments with further analysis in section \ref{sec:V}.   Section  \ref{sec:VI} provides a discussion and the concluding remarks.

\section{Background}
\label{sec:II}
In the current section, we describe the transaction confirmation process, the statement of the problem, and the techniques that we use for model checking.

\subsection{Transaction Confirmation}
Let us consider the transaction confirmation process in Hyperledger Fabric. We choose a blockchain network based on Fabric, as HLF is one of the most popular blockchain frameworks among enterprise applications \cite{b10}. HLF allows specifying the business logic of transactions through the chaincode. Agreement on the chaincode proceeds among network members, which called organizations, in the channel \cite{b24}. Chaincode has an endorsement policy that specifies a set of peers on the channel that must execute the chaincode and endorse the execution results to consider the transaction as valid in the channel.

To specify organizations that might endorse the transaction, we assign a weight to each organization. Authors discuss that weighted multi-party consensus protocols provide more flexibility \cite{b13} \cite{b15}. HLF allows using weighted members in endorsement policies \cite{b12}. 

In weighted consensus protocols, the result depends on the ratio of weights of organizations that reach the consensus. We formalize the endorsement policy using a threshold value $Weight_{th}$. If the organizations in the channel with the total weight greater than $Weight_{th}$ reach the consensus, then the whole network either accepts or rejects the transaction, depending on the responses of organizations that reached the consensus. 

To define a probability for an organization to agree on a transaction, we refer to the historical data that contains responses of organizations. One can collect statistical data on how each organization agrees on a transaction. Using well-known techniques \cite{b8} \cite{b9} chaincode designers can compute the probability to give a refusal response of each organization. The refusal probability defines the probability that an organization does not accept the transaction. We introduce a vector of refusal responses for organizations. Each value in the vector corresponds to the probability to reject a transaction for a certain organization. 

Endorsement policy does not take into account the probabilistic nature of organizations to reach a consensus. We assign to each organization the probability to give a response with which the organization does not reach a consensus. 

\subsection{Model Checking of Probabilistic Systems}
We need to use a model checking technique for probabilistic systems which do not require expensive computations, as usually, designers shall make a decision fast on deployment chaincode in enterprise applications. Moreover, we need to bound an error of computation to control the accuracy of the verification result. We consider different probabilistic model checking techniques and choose the one that fits the most to our requirements.

Given a structure $M$ and a property $\phi$, the model-checking problem answers whether $M$ satisfies $\phi$, which one can write in the following form: $M \models \phi$. We consider specification in a probabilistic Linear Temporal Logic (pLTL) form. The model-checking problem for pLTL properties follows the same steps verification steps as with LTL formulas: model gets composed with the Buchi automaton representation of the verified property, and then checking that the accepting state is visited infinitely often. In pLTL formula, a probability that the model satisfies specification is equal to the probability of reaching a satisfying state. The model checking problem of probabilistic systems is polynomial in the size of the model but double-exponential in the size of the LTL formula \cite{b18}.

Statistical model checking refers to a series of simulation-based techniques that help to answer two questions: (1) Qualitative: Is the probability that the model satisfies the specification is greater than a certain threshold? and (2) Quantitative: What is the probability of the system to satisfy the specification? \cite{b11}. This technique fits better for proving the robust satisfaction of a quantitative property, as the actual probability of satisfying a given specification needs to be bounded away from the threshold to which it is compared \cite{b19}.  As we also need to check the probability threshold to reach a consensus, we use a statistical model checking approach. 
 
 Statistical inference comes in one of two flavors: (a) hypothesis testing is used to determine the extent to which observations “conform” to a given specification, and (b) estimation is used to determine likely values of parameters based on the assumption that the data is randomly drawn from a specified type of distribution. In our modeling, we use the hypothesis testing approach, as estimation appears to be more empiricist in form. In the estimation technique, we need to base an estimate on the hypothesized model that the probabilistic behavior of a system follows some specific probability distribution. From our statistical data, we can not always argue about the probability distribution. In the current work, we model the endorsement policy confirmation system and verify the probability to agree on a transaction. 
 
 \section{Related Work}
\label{sec:III}
Verification using statistical model checking is well studied and numerous statistical model checking tools are developed \cite{b4}. Recent researches use statistical model checking to verify blockchain systems \cite{b5}, \cite{b6}. In provided works, authors study different probabilistic aspects of blockchain network: peer-to-peer system, involving multiple blockchain participants, DNS connection. To the best of our knowledge, there is a gap in the verification of probabilistic aspects of the consensus protocols, which we cover in the current work. 
 
 \section{Modeling and verification}
\label{sec:IV}
In this section, we describe a verification algorithm that takes as an input parameters of the system and outputs the verification result. Verification result gives an answer to the question: can organizations approve or reject a transaction with a certain probability?
The algorithm consists of two phases: constructing a discrete-time Markov chain (DTMC) model and building specification. 

\subsection{Formal Modeling of Multy-Party Consensus Protocols}
The user provides an information about the system with the following parameters:
\begin{itemize}
  \item set of organizations
  \item weight of each organization
  \item probability of each organization to give refusal response for the transaction. This value we take from the vector of refusal responses that one can construct from a historical dataset.
\end{itemize}

And specification parameters:
\begin{itemize}
  \item weight threshold $Weight_{th}$. If the total weight of organizations that reach consensus is more than $Weight_{th}$, then the whole system reaches the consensus
  \item probability threshold $Probability_{th}$. If the whole system agrees on the transaction with the probability of more than $Probability_{th}$, then we consider that the multi-party protocol is valid.
\end{itemize}

We consider parameters of the system in the following form: \textit{Set({organization, probability, weight})}. Each element of the set is a tuple that contains an organization identifier, the probability to give a refusal response, and the weight of the organization. Parameters of the specification one can provide as two values $Weight_{th}$ and $Probability_{th}$. We build a DTMC model from parameters of the system and pLTL formula from parameters of specification. After that, one can check that the model satisfies the specification using statistical model checking.

In Algorithm 1 we show the creation of a DTMC model. First, we take one organization from the set with a corresponding probability and weight. The first organization that we choose is the root of our DTMC model. Next, we take the second organization and assign it to two successors of the root. Transition to one successor happens with a probability $P$, which equals the probability that the root's response is refusal. The second transition happens with a probability $1-P$, which denotes the probability of the root's fair response. After that, we remove the root and successor organizations from the set and recursively apply the algorithm to each subtree. If we traverse from the root to the state with the transition with the probability $P$, then the root's response is refusal, otherwise, the response is acceptance. Set $nodes$ in Algorithm 1 is a set where we save the nodes of the tree. In the input, this set is empty, in the output it keeps all nodes of the tree. Each node keeps the information about the assigned organization, weight, and probability to give a refusal response of the organization and corresponding predecessor. Also, a node keeps the information about the parent's reply: was it refusal or acceptance. If the reply was not a refusal, then the transition from the predecessor happens with a probability $1-P$, otherwise, with a probability $P$. Thus, having a set of nodes one can build a DTMC model. The order of choosing the organizations to build the tree can be random, as variables of the considered stochastic process are independent.

\begin{algorithm}
\caption{DTMC-creation.}
 \KwData{Set(organization, Pr, W) orgs, root, Set nodes}
 \KwResult{Set of nodes for DTMC}
 
 \uIf{orgs is not empty}{
    pop an organization $nextOrg$ with its probability $Pr$ and weight $W$ from the set\;
    
    // build a subtree with a fair response:
    
    create a node $N_{nextOrg_c}$ with root as a predecessor, parameters $W$, $Pr$, and confirmation predecessor's response\;
    
    add the node $N_{nextOrg_c}$ with a fair parent's response to set nodes\;
    
    //run the algorithm recursively to build a subtree for the node with a confirmation //predecessor's response :
    
    DTMC-Creation(orgs, $N_{nextOrg_c}$, nodes)\;
    
    // build a subtree with a refusal response:
    
    create a node $N_{nextOrg_r}$ with root as a predecessor, parameters $W$, $Pr$, and refusal predecessor's response\;
    
    add the $N_{nextOrg_r}$ with a refusal parent's r response to set nodes\;
    
    //run the algorithm recursively to build a subtree for the node with a refusal //predecessor's response:
    DTMC-Creation(orgs, $N_{nextOrg_r}$, nodes)\;
  }
  \Else{
    return nodes \;
  }
 
\end{algorithm}

\subsection{Statistical Model Checking of Consensus Protocols}
Next, we want to check whether the consensus probability $P_{c}$ is greater than the threshold $Probability_{th}$. $P_{c}$ is the probability to get a fair response from organizations with a total weight more than $Weight_{th}$. If the probability to get fair responses from a necessary number of organizations $P_{c}$ is greater than the threshold $Probability_{th}$, then the multi-party protocol is valid.

We assign a value to each node of the DTMC model that corresponds to the total weight of all fair predecessors' weights. We check, that the probability to reach at least one node with the total weight is greater than $Weight_{th}$ is greater than $Probability_{th}$. In other words, we check that the probability to reach the consensus of fair organizations with a total weight more than the weight threshold is greater than the probability threshold. If this property holds, then the model satisfies the specification, and the multi-party protocol is valid. Algorithm 2 takes as an input a DTMC model and outputs nodes with the total weight assignments. The idea is similar to the depth-first search algorithm. But additionally, we assign the total weight to each node. The total weight of the node is a sum of the weight of all predecessors that gave a fair response.

\begin{algorithm}
\caption{labeling-sum}
 \KwData{root, nodes}
 \KwResult{Set of nodes with a total sums assignment}
 
     \ForEach{successor $N_{s}$ of root}{
     assign to $N_{s}$ the total weight of the root\;
     \uIf{root give a fair response }{add the weight of its root to the total weight of $N_{s}$}
       
      labeling-sum($N_{s}$, nodes)\;
    }
 
\end{algorithm}

Once we get the set of nodes with the total weights assignments from Algorithm 2, we can build a pLTL specification. We want to reach at least one of the nodes whose total weight is greater than $Weight_{th}$ with a probability greater than $Probability_{th}$. We can write a disjunction of each node with a total weight greater than $Weight_{th}$. Further, before the disjunction we put operator "Finally" $F$ and the probabilistic operator $P > Probability_{th}$. The final specification tells us that at least one node from the desired set shall be finally reached with a probability greater than $Probability_{th}$.

One can build a specification to compute a transaction rejection instead of confirmation. For this instead of computing the total weight of confirmed responses, one shall compute the total weight of refusal responses and build the specification based on this value.

Let us illustrates the work of two algorithms with an example. We consider the model consisting of three organizations:

\begin{itemize}
  \item Organization 1. Weight 1, Probability 0.07.
  \item Organization 2. Weight 3, Probability 0.01.
  \item Organization 3. Weight 2, Probability 0.02.
\end{itemize}

\begin{figure}[htbp]
\centerline{\includegraphics[width=85mm,scale=0.6]{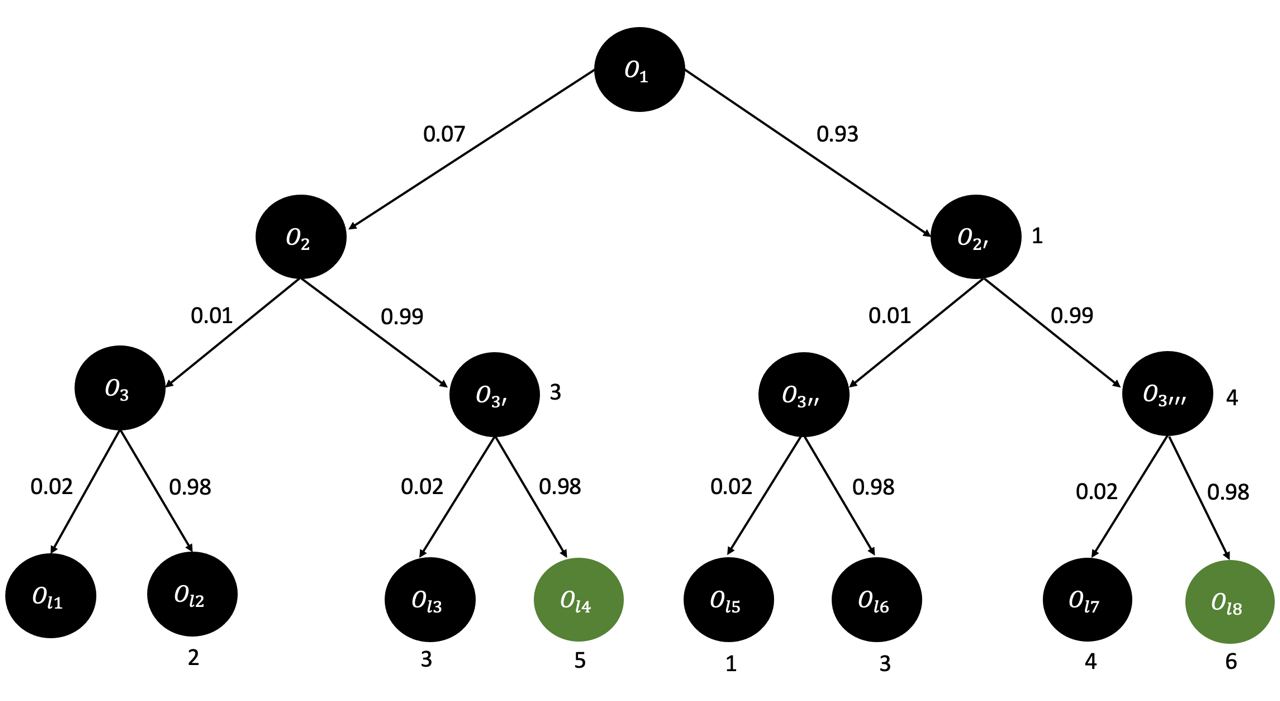}}
\caption{Example of a DTMC endorsement policy model}
\label{figure1}
\end{figure}

We also provide a  specification parameters: $Probability_{th}$ = 0.95, $Weight{th}$ = 5.
Applying Algorithm 1, one can get a DTMC model from  Fig.~\ref{figure1}. Node $O_1$ corresponds to organization 1, nodes $O_2$, $O_2'$ correspond to organization 2, and nodes $O_3$, $O_{3'}$, $O_{3''}$, $O_{3''}$ correspond to organization 3. Nodes $O_{ln}$, where n takes values from 1 to 8, are the leafs of DTMC model. Near from each node we put the total sum that corresponds to this node. We do not write the total sum 0.

Using Algorithm 2 we assign each node its total weight. In Fig.~\ref{figure1} we depict the total weight near each node besides those which total weight is zero. Further, we choose nodes with a total weight greater than $Weight{th}$ to build a pLTL formula. We color these nodes green. The corresponding pLTL specification one can write in the following form:
\[P > 0.95 \;  F \;(O_{l4}\; \;OR \;O_{l8}) \]

With statistical model checking techniques, one can simulate the model. Using hypothesis testing we get the threshold probability when the model satisfies the specification. Our model implies the following assumptions. First, the considered multi-party consensus protocol is ergodistic. We do not consider changing the probabilities to give a refusal reply during the verification process. Second, as we take organization for building the DTMC model in an arbitrary order, we assume that the probabilities of different organizations are independent.

One can use our algorithm to build consensus protocol models with infinite paths. For example, if the threshold weight is not reached with a certain probability on the model, we can make a transition from this leaf to the root. That problem refers to the optimization task. Initially, we need to introduce a metric, the expected number of messages that pass through the system before accepting the transaction. Then, based on this optimization function we can change the model by adding transitions in such a way, that the model satisfies the specification with a minimum expected number of messages. The number of messages is important, as each message can require a fee from the organization that sends it. This problem is interesting for future research.

\section{Experimental Evaluation}
\label{sec:V}
In the current section, we build models and investigate how to change the parameters of the system to satisfy the specification. The goals of the experiments are the following. First, we want to demonstrate the work of our algorithms, and check that the model satisfies invariants of the system. Second, we want to show that relying on the experimental results one can modify the system to make it satisfiable the specifications.

Initially, we provide parameters of the system. We need to choose the number of organizations in our experiments. In the endorsement policy protocol, the number of organizations can be big, i.e. dozens and even hundreds.  But some organizations can be clients of other ones, which will be responsible for the reply on the transaction confirmation proposal. For our experiments, we take 3 organizations, similar to the recent research \cite{b1}. 

Next, we choose weight parameters for each organization in such a way that one organization can not make a decision alone. Also, we take small probabilities to reject a transaction for each organization, i.e. less than 0.1. If the probability to reject a transaction is big, then either there were too many incorrect transactions in the past, or the organization is in an incorrect state, i.e. the organization is not synchronized with the network. Thus, we assume that organizations of our system mostly agree with a transaction. We consider the model similar to one illustrated in Fig.~\ref{figure1}. 

Further, we choose a statistical model checking tool to verify the model. Based on the survey \cite{b4} we have chosen the tool for our experiments that satisfy the following parameters:
\begin{itemize}
  \item based on the hypothesis testing approach
  \item allows to build a DTMC model and check pLTL specifications
  \item simulates the system at the level of the language (on-the-fly), which do not require a memory overhead
  \item the tool is maintained 
\end{itemize}

The first two points correspond to the requirements of our algorithms. The third and fourth points facilitate easy integration and usage of the verification algorithms to Hyperledger Fabric enterprise applications. PRISM tool \cite{b2} satisfy all considered properties. We use PRISM to build a model as a DTMC and a specification in a pLTL form. 

First, we compare verification results based on our algorithms with analytical computations. In the second experiment, we check that the model correctly reflects changes in the system. Finally, we investigate probability versus weight threshold dependency. That shows us how to change the threshold parameters to make the system satisfiable the specification. One can provide considered experiments for the verification of HLF endorsement policies.

We took 10000 samples for each simulation and 0.001 accuracy parameter in PRISM. We run experiments on a 2,3 GHz Dual-Core Intel Core i5 CPU. The memory limit is 2 GB. We did not put the time limits, as each experiment takes less than 1 second.

\subsection{Experimental Evaluation of Verification Algorithms}
Initially, we want to prove with experiments that our algorithms for building models and specifications are valid. We claim that the proposed method to build a model corresponds to the theoretical computations of probability to accept a transaction. 

 The probability threshold for our models is $0.95$, the weight threshold is 5. Fig.~\ref{figure1} illustrates the model for the experiment. The code of the models is available in the Github repository \cite{b26}. We take examples that allow computing analytically the probability to accept a transaction.  We considered different cases:
\begin{itemize}
  \item All organizations give the same response on the transaction. 
  \item Organizations 2 and 3 with weights 3 and 2 correspondingly agree on the transaction with probability 1, organization 1 refuses a transaction with a probability 1.
  \item Organization 1 with weight 1 agrees on the transaction with probability 1, other organizations refuse a transaction with probability 1.
  \item All organizations agree on the transaction with a probability of 0.5.
\end{itemize}

We build models for each case and compute the probability that the transaction gets accepted. As the weight threshold is 5, desired nodes correspond to states $O_{l4}$ and $O_{l8}$ from the model in Fig.~\ref{figure1}. 

As expected, in the first case PRISM returns the probability 1 to reach a consensus. The second experiment also shows probability 1. That corresponds to the analytical computation, as organizations with a total weight greater than the threshold accept a transaction with a probability of 1. In the third case, the probability to accept a transaction is 0, as the majority of organizations give a refusal response with a probability of 1. The experiment shows the same result. In the fourth experiment, the probability to accept a transaction is 0.25. We get the same result analytically by multiplying probabilities. 

\subsection{Effect of the Changing System Parameters on the Verification}
In this section, we check that changing the system parameters correspond to the changes in the model and verification result. That proves that the system corresponds to the constructed model. We considered the following scenarios:
\begin{itemize}
  \item Invariant 1. The probability to accept transactions decreases for one organization. We expect that the overall acceptance probability shall decrease as well.
  \item Invariant 2. One organization leaves the channel. The probability of the whole system to accept the transaction shall either remain the same or decrease.
\end{itemize}

We take the model from Fig.~\ref{figure1} with the same probability and weight thresholds, $0.95$ and $5$ correspondently.

\subsubsection{Invariant 1}
In the current scenario, we decrease the probability to accept the transaction for organization $O_3$. We provide the dependency of the probability to accept the transaction for organization 3 versus the probability that the whole system agrees on the transaction. We decrease the probability to accept a transaction for an organization iteratively on 0.01. 

\begin{figure}[htbp]
\centerline{\includegraphics[width=85mm,scale=0.6]{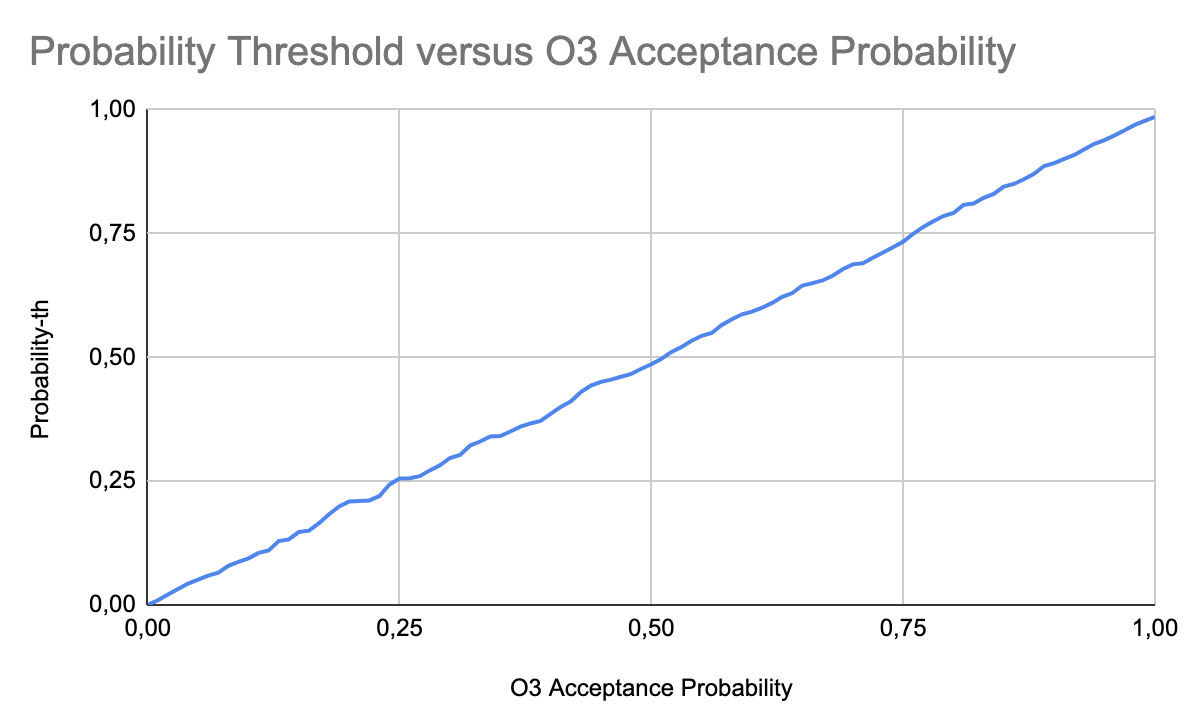}}
\caption{Probability Threshold versus O3 Acceptance Probability Dependency}
\label{figure2}
\end{figure}

We illustrated the experimental result in Fig.~\ref{figure2}. The dependency is expected: with decreasing of acceptance rate of organization 3, the probability threshold of the whole system to accept the transaction decreases as well. We see that the dependency is linear, and the discrepancy is less than $0.001$, which corresponds to the experiment setup. 

\subsubsection{Invariant 2}
\begin{figure}[htbp]
\centerline{\includegraphics[width=85mm,scale=0.6]{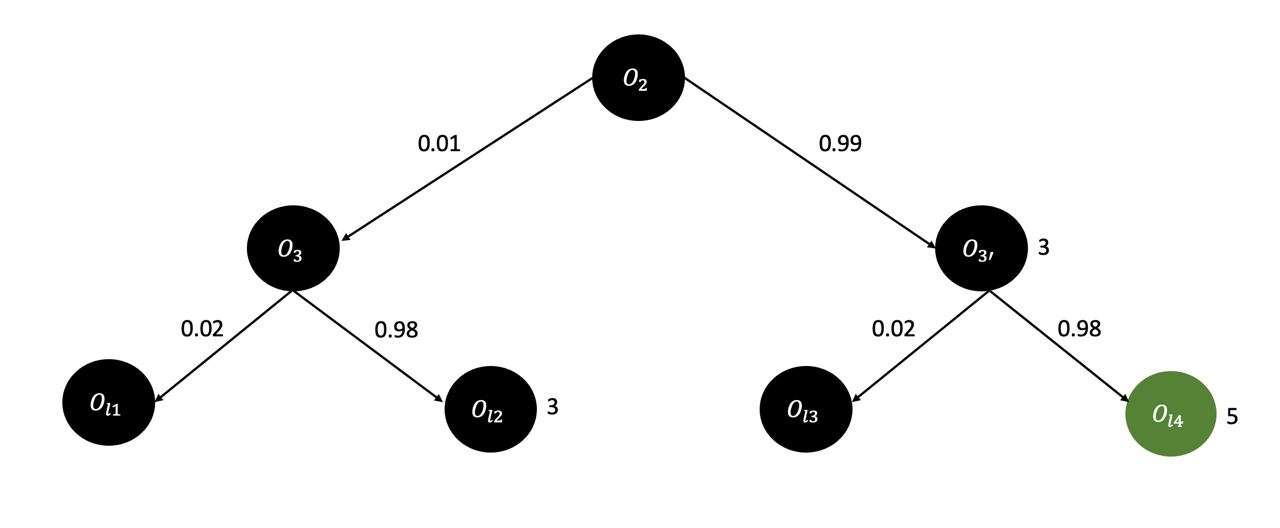}}
\caption{DTMC model with 2 organizations}
\label{figure3}
\end{figure}

Now we investigate how changing the number of organizations in the channel affects the probability to accept the transaction by the whole system. We remove organization 1, as the total sum of organizations 2 and 3 still equals the threshold. Applying algorithms 1 and 2, we can get the new model and nodes on this model reaching which the whole system accepts the transaction. Fig.~\ref{figure3} illustrates this model, the green node is a node returned by Algorithm 2.

Simulating the model we get that the probability of the whole system to accept the transaction is $0.9683$. For the system with three organizations the probability to accept the transaction is $0.9689$. We see that the difference between acceptance probability is less than the discrepancy in our experiment, which is $0.001$. Thus, removal organization 1 does not affect the probability of the whole system accepting the transaction. This is the expected result, as the total sum of organization 1 with either organization 2 or organization 3 is less than the weight threshold of $5$. Hence, the reply of organization 1 does not affect the probability of the system accepting the transaction. We also can see it in Fig.~\ref{figure1}: the left and the right subtree of the node $O_1$ are the same and similar to the tree from Fig.~\ref{figure3}.

\subsection{Relation of the Specification Parameters}

In this experiment, we get the dependency of the weight threshold versus the probability threshold. Table 1 illustrates the dependency.  The result is expected: if the weight threshold is low, then the probability to accept a transaction is also low, as in our model acceptance probability of each organization is high. With an increase of the weight threshold, the acceptance probability decreases. When the weight threshold is greater than 6, then the probability to refuse transaction will be constant 0, as there are no organizations in our model with a total weight greater than 6. 

\begin{table}[h]
\caption{Weight threshold versus probability threshold dependency }
\begin{tabular}{|c|c|}
\hline
Weight threshold & Probability threshold \\
\hline
1 & 0.9990 +/-0.0004 \\
\hline
2 & 0.9997 +/-0.0005 \\
\hline
3 & 0.9989 +/- 0.001 \\
\hline
4 & 0.9894 +/- 0.0003 \\
\hline
5 & 0.9717 +/- 0.0005 \\
\hline
6 & 0.9035 +/- 0.0009 \\
\hline
7 & 0 \\
\hline
\end{tabular}
\end{table}

\subsection{Analysis of the Experimental Results}
We proved with experiments that our algorithms for building models and specifications are valid and invariants of the system remain the same.  Although we cannot provide experimental evaluation for all possible cases, one can see that the experimental results based on our algorithms correspond to the analytical computations. We also have shown that if one changes the system parameters, then the model and verification results also change in the right direction. Based on provided dependencies, one can choose probability and weight thresholds in such a way that the model satisfies the specification.

\section{Conclusion and Future Work}
\label{sec:VI}
In this paper, we analyzed the multi-party consensus problem. To check that the protocol satisfies the specification we proposed algorithms to build a DTMC model and a pLTL specification. Further, we verified the model using the statistical model checking approach. One can use proposed algorithms and experiments to verify endorsement policies in the Hyperledger Fabric framework, as it allows to assign weights to organizations. 

As further research,  we want to investigate the scheduling of sending the confirmation messages on the model with infinite paths. Also, we are going to collect the data of real-world endorsement policies to provide experiments on them.

\bibliographystyle{ieeetr}
\bibliography{bibliography} 

\end{document}